# Turbulence-induced oscillation on particle detachment from a wall

Zhikai You[1], Yiyang Zhang[1,*], Zhu Fang[2] and Shuiqing Li[2]

[1]*Institute of Nuclear and New Energy Technology, Tsinghua University, Beijing 100084, China*

[2]*Department of Energy and Power Engineering, Tsinghua University, Beijing 100084, China*

*\*zhangyiyang@mail.tsinghua.edu.cn*

Particle resuspension is a ubiquitous phenomenon with pivotal relevance in numerous natural and industrial contexts. In this study, we present findings on the resuspension of individual micro-sized particles, captured through high-speed camera experiments. Our observations reveal a universal behavior whereby a particle undergoes oscillatory motion due to turbulent excitation prior to its detachment from the surface. This motion is characterized by dimensionless number *Ad* and *S*. The frequency of particles oscillation is analyzed and it shows the frequency of particle oscillations increased with decreasing particle size. We establish a new model that the particle is a linear oscillator driven by stochastic torque from turbulence. It is shown that the stochastic oscillation is the key mechanism for particle detachment from a wall within a certain range of friction velocities.

Small particles resuspension phenomena exist widely in nature environments and industry applications, such as the resuspension of radioactive dust in nuclear reactors (including fission reactors and fusion reactors) [1-3], dust removal from the surface of spacecraft solar panels on Mars [4,5], and particle sampling and detection [6,7]. In recent years, surface cleaning techniques for particle contamination control have become increasingly important in several engineering fields, notably in the semiconductor manufacturing industry [8-10]. Despite nearly five decades of study on particle resuspension in turbulent boundary layers, there is still no unified theory on the mechanism of particle detachment from a wall.

Small particles deposited on a wall are bound to the surface by strong adhesive forces, which are usually the result of chemical bonding coupled with mechanical stress. When turbulence flow passes over the surface, it breaks this adhesive contact and entraps the particles. At present, models for particle resuspension or detachment from the wall are categorized into two types: the force/moment balance model and energy accumulation models which are well summarized in the review [11,12]. The former believes that the particles separate from the surface because they are subjected to an aerodynamic moment that surpasses the adhesion moment. The energy accumulation model, however, introduces the concept of adhesion energy potential well. The most famous of the energy accumulation models are the RNR model [13] and VZFG model [14], which describe the linear and nonlinear oscillations of particles on the surface induced by turbulence, respectively. Nevertheless, the use of these models relies on the fitting of parameters and these models still lack experimental evidence. So far, there have been also some experiments studies. By observation, the motion of the particles during resuspension is divided into rolling, lifting and bouncing [15]. Additionally, a positive correlation between the resuspension rate of particles and the local turbulence intensity of the airflow has been documented under varying flow conditions experiments [16,17]. In a number of experiments using air jet to remove particles from surface, it has been found that introducing velocity fluctuations to the flow field significantly increased particle removal [8,18].This implies that particle resuspension events are related to the fluctuating component of the turbulence. However, how particles detach from the wall on smaller temporal and spatial scales remains elusive.

In present work, we employ high-speed

photomicrography to investigate the behavior of a single micron-sized spherical particle on a smooth surface during resuspension within a channel. We find the key experimental evidence of particle oscillation induced by turbulence. We investigate the relationship between the frequency of the particles and their size and discover this oscillation pattern is inconsistent with that described by RNR and VZFG models. Thus, we derive and establish a new model of the oscillation behavior based on the classical theory of adhesive contacts. The experimental results are in general agreement with those obtained from the model predictions.

The experiment is conducted within a rectangular channel (10×10×1500 mm), equipped with a flow controller to ensure precise regulation of air flow. Silica and alumina spheres with particle sizes of 50-110 μm are deposited on a quartz substrate (roughness less than 0.1 nm) in the test section after drying. A high-speed camera, outfitted with a microscope head is used to record the resuspension of individual particles (see sketch in Fig. 1) and its frame rate is 8600 fps. The magnification of the microscope head is 28x. The channel is long enough that turbulence is fully developed in the test section. The experiment commences by adjusting the camera's focal plane to a single particle, followed by a gradual increase in the channel's airflow velocity from zero to avoid acceleration effects. The surface energy of the particles is greatly reduced after the surface has been treated with a silane coupling agent. By measuring, the effective surface energy γ between the particles and the quartz substrate are 0.034 mJ/m$^2$ (γ-SiO$_2$) and 0.092 mJ/m$^2$ (γ-Al$_3$O$_2$).

As the reported experiments [11], we also observed rolling and lifting motion of the particles. Moreover, we have discovered the fact that prior to detachment, the particles exhibited oscillatory behavior when viewed from both the side and top view, as depicted in Fig.2. This may be due to the small size of the particles and the weak velocity fluctuations in the spanwise direction. Videos of the experiments are provided as supplementary material. Our observations reveals that with a gradual increase in airflow velocity, the particle transitions through distinct states of motion: static, oscillating, and ultimately rolling or lifting, which suggests the described phenomenon is ubiquitous. After binarization and filtering the recorded images, we are able to calculate the form-center coordinate and thus obtain the displacement information of the micro-spheres. Analysis of the entire video recording yielded the displacement-time series data for a particle oscillating on the surface. For instance, the results of a 62 μm SiO$_2$ sphere from one experiment, presented in Fig. 3, indicate that the oscillation duration was approximately 0.15 seconds, which is relatively brief. The oscillation frequency of the particle can be obtained by Fast Fourier Transform (FFT). For this particle, the oscillation frequency is about 1321 Hz. This value is reliable because Nyquist theorem tells us that values with frequencies less than half the frame rate are plausible.

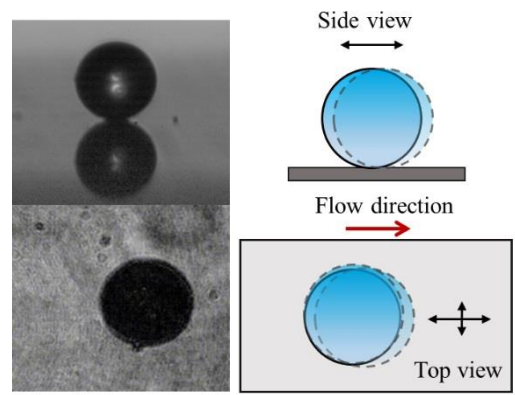

FIG.2 High-speed observations of a single particle motion before resuspension on (a) side view and (b) top view.

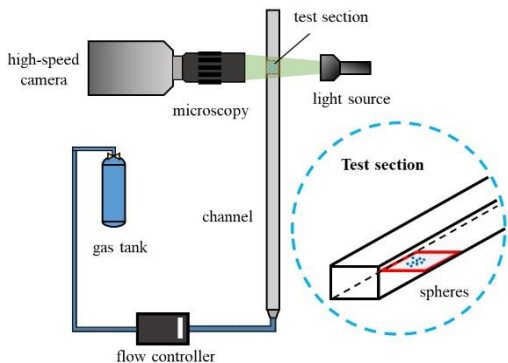

FIG.1 Sketch of the experimental setup.

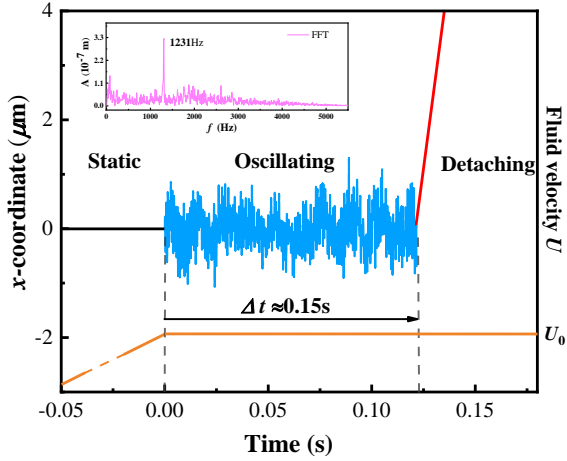

FIG.3 Displacement-time series of a 62 μm SiO$_2$ sphere.

It seems that oscillatory motion of particles occurs only under certain conditions. In the following, we present a theoretical analysis of this phenomenon. The consensus is that particle resuspension is the result of coupled hydrodynamics and particle adhesion. The adhesion force between a particle and the surface which is regarded as acting at the center of the contact circle is related to the surface energy $\gamma$ of the material and particle size $R$. Based on JKR model [19], the adhesion force $F_c$ is given by:

$$F_c = \frac{3}{2}\pi\gamma R. \quad (1)$$

The radius of the contact circle $a_0$ is

$$a_0 = 1.26(3\pi\gamma R^2/k)^{1/3}, \quad (2)$$

where $k$ denotes the elastic constant,

$$k = \frac{4}{3}\left(\frac{1-\nu_1^2}{E_1} + \frac{1-\nu_2^2}{E_2}\right)^{-1}. \quad (3)$$

Consequently, the moment caused by adhesion force can be estimated as,

$$T_{ad} = \frac{3}{2}\pi\gamma R \times 1.26(3\pi\gamma R^2/k)^{1/3}. \quad (4)$$

The average drag force on a sphere in the predominantly viscous sub-layer can be expressed as [20]

$$\frac{\langle F_d \rangle}{\rho\nu^2} \approx 32\left(\frac{Ru_\tau}{\nu}\right)^2, \quad (5)$$

where $u_\tau$ is friction velocity and $\nu$ is the kinematic viscosity. Generally, the moment caused by lift force on the particle is significantly smaller than that caused by the drag force ($a_0 \ll R$), thus, the contribution of the lift force can be ignored. By comparing the average drag moment with the recovery moment, we can define a dimensionless number, $\beta$, as:

$$\beta = 2.55 \frac{\rho_f u_\tau^2 R^{4/3} k^{1/3}}{\gamma^{4/3}}. \quad (6)$$

It can be deduced that a particle will detach from the surface by either rolling or lifting when $\beta > 1$. Conversely, when $\beta < 1$, it is uncertain whether the particle will break away from the surface, as it is also related to local turbulence. The particles may also re-suspend by oscillation and energy accumulation. Further, Eq.6 can be rewritten as follows,

$$\frac{\gamma}{\rho u_\tau^2 R} = 2.55\frac{1}{\beta}\left(\frac{Rk}{\gamma}\right)^{1/3}. \quad (7)$$

Here, we obtain two dimensionless numbers, the adhesion number $Ad$ and the dimensionless stiffness $S$.

$$Ad = \frac{\gamma}{\rho u_\tau^2 R} \quad (8)$$

$$S = \frac{Rk}{\gamma} \quad (9)$$

The adhesion number $Ad$ represents the ratio of the surface energy of particles to the kinetic energy transferred by the turbulence and the dimensionless stiffness $S$ represents the ratio of the stored elastic potential energy to its surface energy. This implies that the resuspension event is the result of competition among the turbulent energy, surface energy and the elastic potential energy. These parameters are utilized to construct a phase diagram, and experimental results are incorporated into this diagram, as shown in Fig. 4. The motion states of the particles are classified as static, oscillating and rolling/lifting.

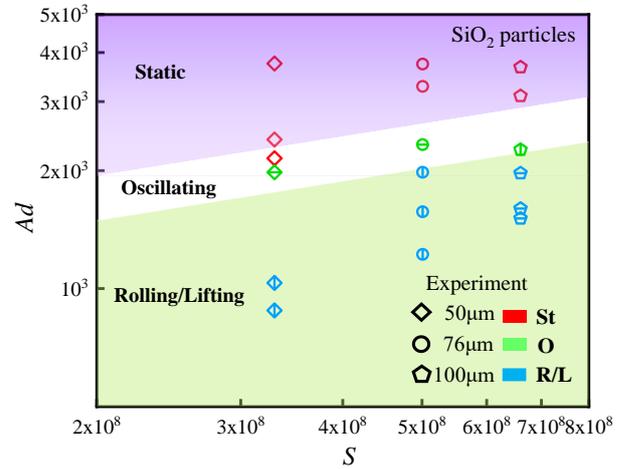

FIG.4 Regime map of particle motion before detachment.

Beyond the aforementioned discussion, our primary interest lies in the oscillation frequency of the particles, which serves as an indicator of the mechanisms underlying particle resuspension. We recorded the frequency of a number of silica and alumina spheres of different sizes as they oscillate. The following conclusions are obtained by analyzing these results.

(i) *The frequencies observed in the experiments exhibit a degree of randomness.* Since the surface of the particles is not absolutely smooth and the particles are not perfectly spherical, the conditions of adhesion of each particle in contact with the wall are different.

(ii) *The frequency of particles increases as particle size decrease.* Smaller particles required greater airflow velocities for resuspension, and are also observed to oscillate at a greater frequency (see the result in Fig. 6).

In previous models, the adhesive contact between the particle and the wall is described as a linear or nonlinear spring. The stiffness of this spring is derived using the Johnson-Kendall-Roberts (JKR) model by Reeks *et al.*, and the detailed derivation is available in references [13,21]. The expression for this stiffness between particles and surface is adopted in both RNR and VZFG model,

$$\chi = \frac{dP}{d\delta} = \frac{9}{2} R^{1/3} \kappa^{2/3} P_1^{1/3} \frac{P_1+P}{5P_1+P}. \tag{10}$$

Where $P$ is a normal load, $\delta$ is the overlap, and $P_1$ is the effective load, $P1=P+3\pi\gamma R+[3\pi\gamma RP+(3\pi\gamma R)^2]^{1/2}$. In quasi-static experiments, the particle eventually pull off at the adhesion force [Eq. (1)] as the load $P$ increases. For a 50µm $SiO_2$ sphere, the frequency $f = (\chi/m)^{0.5}$ is about $10^5$ Hz, which is much higher than our experiments. In addition, in our experiments, we observed that the particles preferred a rotational motion. This makes us believe that the particle is deformed in a rotational style to produce a restoring torque, rather than a normal deformation.

In present study, we obtain an expression for the particle rotation torque from the classical theory of adhesion contact. When no turbulence is present, the radius of the contact is $a_0$. In JKR framework, $a_0= (9\pi\gamma R^2/E)^{1/3}$. The particle first slightly drifts from its original position at a small angle $\theta$ by torque from turbulence. Following Dominik's work [22], we assume that the shape of the contact area becomes two semicircles with radius $a_0+\xi$ and $a_0-\xi$, respectively. This is not inconsistent with Savkoor's [23,24] analysis of tangential behavior in the presence of adhesion conditions and it is this asymmetry that produces the restoring torque. The schematic is shown in Fig.5.

The oscillation is driven by stochastic aerodynamic torques.

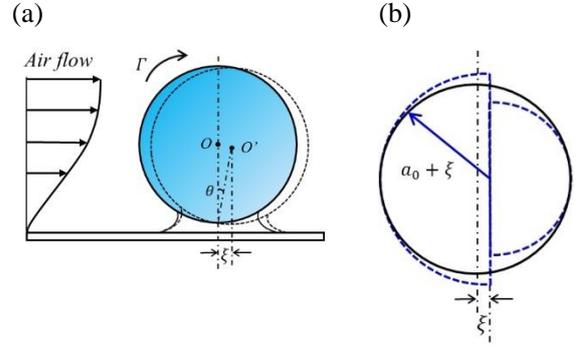

FIG.5 Schematic of particle deformation in the presence of torque: (a) rotation (b) contact area.

The rolling resistance moment is given by Dominik [22], by solving the Boussineq problem,

$$M_y = 4\pi F_c (a/a_0)^{3/2} \xi. \tag{11}$$

In this case, the factor $a/a_0$ is approximately 1. Substituting Eq. (1) into Eq. (11) and considering $\xi \approx R\theta$, we have

$$M_y \approx 6\pi\gamma R^2 \theta. \tag{12}$$

Thus, the motion of particle can be described as,

$$I\frac{d^2\theta}{dt^2} + 6\pi\gamma R^2 \theta = \Gamma(t), \tag{13}$$

where $\Gamma$ is the stochastic torque induced by turbulence and $I$ is moment of inertia for the particle. For this linear oscillating system, we can obtain its natural frequency $f_0$,

$$f_0 = \sqrt{\frac{45}{14} \frac{\gamma}{\rho_s R^3}}, \tag{14}$$

where $\rho_s$ is density of the particle. The natural frequency $f_0$ of the system has a -1.5 power law relationship with the radius of the particles, which implies that microspheres with smaller inertia have higher frequencies. This result is similar to that of Peri *et al*. [25] for particles under acoustic excitation. Our experiments verify this conclusion and the results are exhibited in Fig. 6.

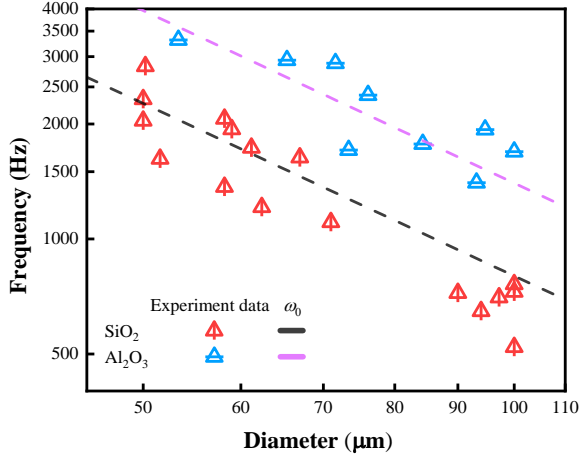

FIG.6 Frequency of particles with different sizes. (Symbols: frequency obtained from experiments; dashed lines: natural frequency predicted from Eq. (14)).

Once the deformation of the particles exceeds a certain threshold $\theta_{crit}$, the particle will detach from the surface and the oscillator system is no longer existent. However, the parameter $\theta_{crit}$ is related to surface morphology and experiment environment, which is supported by Farzi's experiments [26]. Because of this, some randomness is shown in our experimental results.

Here we revisit the physical facts of the particle detachment or resuspension phenomenon: the fluid exerts a torque on the particle, and the restoring torque is provided by an asymmetric pressure distribution in the contact region. The particles accumulate energy by oscillating in turbulence. If we consider the separation of particles as a kind of "damage" to the contact structure, then this phenomenon can be described by stochastic damage theory. In particular, the system suffers damage when the response $\theta$ exceeds an upper limit $\theta_{crit}$ for the first time. In the absence of significant changes in friction velocity, resuspension can be regarded as a smooth stochastic process. Mollinger et al. [27] have measured the lift force of a small sphere in a viscous sublayer, which turned out to be Gaussian. Since the flow structures that induce lift force and drag force are the same, we consider the torque to be Gaussian as well. On this assumption, the expectation of the frequency that spans $\theta = \theta_{crit}$ with a positive slope $\nu_a^+$, can be expressed as,

$$\nu_a^+ = \int_0^\infty \dot{\theta} p(\theta_{crit}, \dot{\theta}) d\dot{\theta}, \quad (15)$$

where $p(\theta_{crit}, \dot{\theta})$ is the value of the 2D probability density function $p(\theta, \dot{\theta})$ at $\theta = \theta_{crit}$. We assume that $\theta$ and $\dot{\theta}$ follow a joint normal distribution such that the expectation of the damage frequency can be expressed as,

$$\nu_a^+ = \frac{\sigma_{\dot{\theta}}}{2\pi\sigma_\theta} e^{-\frac{\theta_{crit}^2}{2\sigma_\theta^2}}, \quad (16)$$

The coefficient $\sigma_{\dot{\theta}}/2\pi\sigma_\theta$ in Eq. (16) represents the natural frequency $f_0$ of the particle oscillating system, i.e.

$$\nu_a^+ = f_0 e^{-\frac{\theta_{crit}^2}{2\sigma_\theta^2}}. \quad (17)$$

Thus, the probability $P$ that a spherical particle would detach from the wall under turbulent excitation at time $\Delta t$ is $\nu_a^+ \cdot \Delta t$. When the effect of particle inertia is neglected, $\sigma_\theta$ can be obtained by the impulse response function method as,

$$\nu_a^+ = f_0 \exp\left[-\frac{(M_{crit} - \langle\Gamma\rangle)^2}{2\langle\zeta^2\rangle}\right], \quad (18)$$

where $\zeta$ is the fluctuation torques and $M_{crit}$ is critical torque. This conclusion is similar to that given by Reeks et al. [20] in analogy with chemisorption and dissociation processes.

The probability of particle detachment depends on the turbulence parameters and the natural frequency of the particle. For a 1μm $SiO_2$ sphere, the frequency is about $4\times10^4$ Hz (in RNR model, it is about 10MHz). This means those difficult-to-remove micro-sized particles may be able to be removed by means of resonance. On the other hand, we also observed oscillations of particles on substrates with significant roughness. Rough surfaces seem to tend to make the oscillation frequency lower. This will be discussed in future work.

In summary, we have found the key evidence of turbulence-induced particle oscillations and resuspension through high-speed camera experiments. The motion of the particles before detaching from the wall can be divided into static, oscillating and rolling or lifting, which is controlled by two dimensionless numbers, the adhesion number $Ad$ and the dimensionless stiffness $S$. Particle oscillations can be described as a linear oscillator driven by stochastic torque $\Gamma$ whose natural frequency $f_0$ is related to the density, surface energy and size of the particle. Thus, our finding may advance the understanding on micro-particles resuspension phenomena in natural environments and industrial process. More importantly, it may be able to provide instructions for the design of cleaning technologies for particle contamination in special situations, making it possible to utilize resonance for particle removal.